\documentclass[pre,twocolumn,aps,eqsecnum]{revtex4}
\usepackage{graphicx}

\begin{document}

\title{Thermodynamic analysis of the Livermore molecular-dynamics simulations of dislocation-mediated plasticity} 
\author{J.S. Langer}
\affiliation{Department of Physics, University of California, Santa Barbara, CA  93106-9530}
\date{\today}

\begin{abstract}
Results of recent large-scale molecular dynamics simulations of dislocation-mediated solid plasticity are campared with predictions of the statistical thermodynamic theory of these phenomena.  These computational and theoretical  analyses are in substantial agreement with each other in both their descriptions of strain-rate dependent steady plastic flow and of a transient stress peak associated with initially small densities of dislocations.  The comparisons between the numerical simulations and basic theory  reveal inconsistencies in some conventional phenomenological descriptions of solid plasticity. 
\end{abstract}
\maketitle

\section{Introduction}

The Livermore molecular dynamics simulations (LMD) \cite{Bulatov-17} of dislocation-mediated solid plasticity are the first numerical computations that are large enough to produce realistic descriptions of this centrally important class of materials phenomena.  Similarly, the thermodynamic dislocation theory \cite{LBL-10,JSL-17a} has been the first apparently successful attempt to develop a dislocation theory based on the principles of nonequilibrium statistical thermodynamics, and to use that theory to interpret a wide range of experimental observations.\cite{JSL-15,JSL-15y,JSL-16,JSL-17,LTL-17,LTL-18}   My purpose here is to show that these computational and theoretical  approaches are in substantial agreement with each other.  Moreover, the process of comparing these approaches reveals  inconsistencies in some of the phenomenological assumptions that are common in this field. 

The LMD results do not suffer from the intrinsic limitations of discrete dislocation dynamics \cite{KUBIN-08}, which, so far as I know, have not made realistic contact with phenomena such as strain hardening or yielding transitions.\cite{LESAR-14}  In contrast, the LMD investigators have simulated three dimensional systems containing $10^9 - 10^{10}$ tantalum atoms interacting via realistic embedded-atom potentials. Using one of the world's most powerful computers, they have been able to drive their systems at strain rates as low as $10^5 \,s^{-1}$, and thus are beginning to make contact with laboratory experiments.  In fact, these simulations are better than real experiments in the sense that temperatures and initial conditions are controlled, and that the density of dislocations as well as the stress is measured directly. 

I focus here on two aspects of these numerical simulations. First, they confirm that simulated crystalline systems can be driven into steady states of plastic deformation at room temperature and at strain rates up to about  $10^8\,s^{-1}$.  This is a regime in which the thermodynamic theory is based directly on fundamental principles. It argues, for reasons that I think are compelling for crystals with dislocations, that there must be an effective temperature $T_{eff}$ that measures the degree of configurational disorder.  The theory then assumes for kinematic reasons that $T_{eff}$ is determined solely by the rate of plastic deformation, i.e. by the rate at which the system is being ``stirred'' by the external driving.  In turn, $T_{eff}$  determines the steady-state density of dislocations and thus the stress required to drive the deformation.  The simulations and the theory are in quantitative agreement in this regime, which lends credibility to both approaches.  

Second, and just as interesting, is that the systems modeled in the LMD simulations are seeded with small initial densities of dislocations and, as a result, exhibit peaks in their early-stage stress-strain curves.  These peaks look qualitatively different from those seen in laboratory experiments or in earlier theories.  In the laboratory, samples prepared for high-strain-rate testing generally contain appreciably large densities of dislocations.  After brief periods of elastic deformation, these experimental samples undergo sharp elastic-to-plastic yielding transitions followed by stages of strain hardening and then steady flow.  In the LMD simulations, however, the initial elastic deformation persists to larger stresses, and a yielding transition is followed by a marked drop in the stress, i.e. a transient softening stage rather than a hardening one.  The peak is accompanied by a rapid increase in the density of dislocations, which then accounts for more familiar stress-strain behavior after the peak has relaxed.  

My explanation for this behavior is that the initial dislocation segments are stretched and accelerated by the increasing stress via a drag force in which the normal velocity of a dislocation line is  proportional to the stress.  This is an irreversible process in which some fraction of the power delivered to the system is converted into the energy of new dislocations.  At first, these new dislocations are not dense enough to become strongly entangled with each other; thus, their appearance is a softening mechanism.  They enable deformation, and the stress drops.  As their density increases, they become entangled (e.g. see \cite{Bulatov et al - 06}), and the system undergoes conventional strain hardening.

The crossover between these dislocation mechanisms is not trivial.  The drag stresses and entanglement stresses cannot simply be added to one another, as is often done in phenomenological analyses.  For example, see Eq.(1) in Gray's recent review article\cite{GRAY-12}, where the stresses associated with a variety of impediments to dislocation motion are assumed to contribute additively to the total flow stress.  Or see Eq.(1) in Armstrong's review of Hall-Petch effects \cite{ARMSTRONG-16}, where the author assumes that the stress associated with dislocation pile-ups at grain boundaries can simply be added to the drag stress that moves dislocations across the interiors of the grains.   

The LMD simulations provide a perfect opportunity to examine a situation in which two different impediments to dislocation motion -- drag and entanglement -- are acting simultaneously, in close analogy to the situations addressed in the conventional literature.  There is no uncertainty about the equations of motion for this situation, which are presented here in Sec.\ref{stress-peak}.  The stresses do not add in the conventional way.  On the contrary, the equations predict a highly nonlinear crossover between drag control and entanglement control, a prediction that is confirmed in detail by the simulations.  I emphasize that this detailed comparison is possible only because the simulations accurately control the initial conditions and accurately monitor both the stress and the dislocation density throughout the process.

\section {Basic Elements of the Thermodynamic Dislocation Theory: The Effective Temperature}

The thermodynamic dislocation theory is applied here to phenomena that are not exactly the same as those in my recent publications. Therefore, I start by restating some basic assumptions.  The next several paragraphs are adapted from Ref.\cite{JSL-17a}, modified to focus on the higher strain rates that are important here.  

In contrast to amorphous plasticity \cite{FL-11}, where identifying shear transformation zones or the like has always been problematic, the elementary flow defects in crystals, {\it i.e.} the dislocations, are unambiguous.  They are easily identifiable line defects, whose energies and dynamic time scales differ from those of the ambient thermal fluctuations by many orders of magnitude.  The dislocations have well defined energies and easily visible configurations.  Thus it is possible to make crude but plausible first estimates of thermodynamic quantities.  

To explore this picture, it is useful to think of a slab of material lying in the plane of an applied shear stress, undergoing only steady-state deformation, and to focus only on the dislocations.  The dislocation lines oriented perpendicular to this plane are driven by the stress to move through a forest of entangling dislocations lying primarily in the plane, thus producing shear flow. Let the area of this slab be $A_0$, and let its thickness be a characteristic dislocation length, say $L_0$.  Denote the configurational energy and entropy of the slab by $U_0(\rho)$ and $S_0(\rho)$ respectively.  Here, $\rho$ is the areal density of dislocations or, equivalently, the total length of dislocation lines per unit volume. The entropy $S_0(\rho)$ is to be computed by counting the number of arrangements of dislocations at fixed values of $U_0$ and $\rho$.  

The dislocations are driven by the applied stress to undergo motions that are chaotic on deformation time scales; that is, they explore  statistically significant parts of their configuration space.  According to Gibbs, this configurational subsystem maximizes its entropy; that is, it finds a state of maximum probability.  It does this at a value of the energy $U_0$ that is determined by the balance between the input power and the rate at which energy is dissipated into a thermal reservoir.  The method of Lagrange multipliers tells us to find this most probable state by maximizing the function $S_0 - (1/\chi)\, U_0$, and then finding the value of the multiplier $1/\chi$ for which $U_0$ has the desired value.  Thus $ \chi \equiv k_B T_{e\!f\!f}$; and the  the free energy to be minimized is
\begin{equation}
\label{Fdef}
F_0 = U_0 - \chi\,S_0. 
\end{equation}

Minimizing $F_0$ in Eq.~(\ref{Fdef}) determines the steady-state dislocation density, say  $\rho_{ss}$, as a function of the steady-state effective temperature $ \chi_{ss}$. In the simplest approximation, $U_0 = A_0\,e_D\, \rho$, where $e_D$ is a characteristic energy of a dislocation of length $L_0$.  (As in previous papers, I omit the conventional logarithmic correction for elastic energy primarily because it muddies the algebra unnecessarily, but also because I am not sure it is correct for present purposes.)   Similarly, we can estimate the $\rho$ dependence of the entropy $S_0$ by dividing the  area $A_0$ into elementary squares of area $a^2$, where $a$ is the minimum spacing between noninteracting dislocations -- an atomic length scale, probably somewhat larger than the length of the Burgers vector $b$.  Then we count the number of ways in which we can distribute $\rho\,A_0$ line-like dislocations, oriented  perpendicular to the plane, among those squares. The result has the familiar form $S_0 = -\,A_0\,\rho\,\ln(a^2\,\rho) + A_0\,\rho$.  Minimizing $F_0$ with respect to $\rho$ produces the usual Boltzmann formula, 
\begin{equation}
\label{rhoss}
\rho_{ss} = {1\over a^2}\,e^{-\,e_D/\chi_{ss}}. 
\end{equation}
We see that an appreciable density of dislocations requires a value of $\chi= \chi_{ss}$ that is comparable to $e_D$, which is enormously larger than the ambient thermal energy $k_B\,T$.   We also see that $a$ is the spacing between dislocations in the limit in which $\chi_{ss} \to \infty$ and thus is an upper bound for the validity of this theory.  

\begin{figure}[h]
\begin{center}
\includegraphics[width=\linewidth] {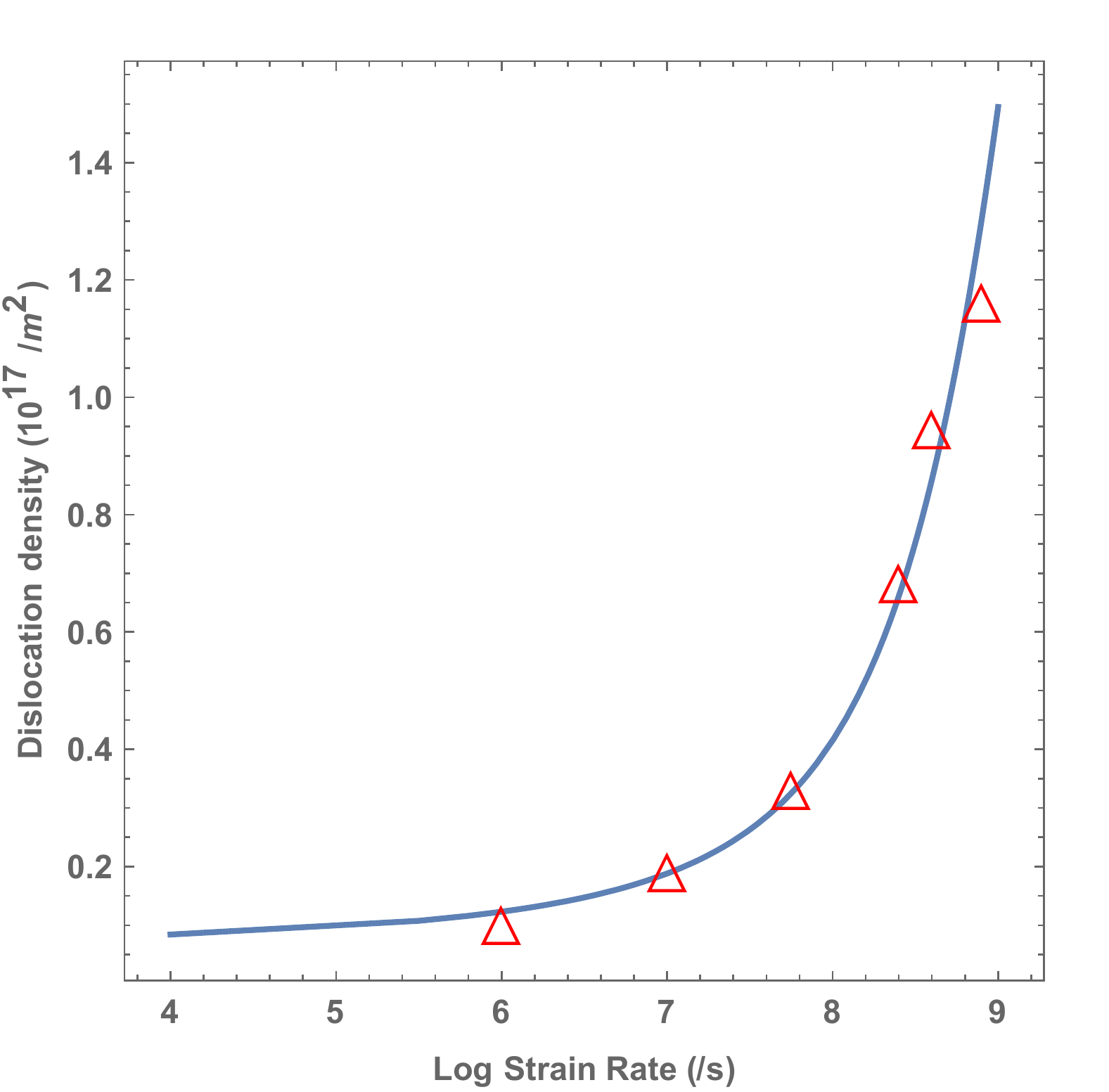}
\caption{Dislocation density as a function of strain rate.  The data points are taken from \cite{Bulatov-17}}
\label{Fig1}
\end{center}
\end{figure} 

Next, note that $\chi$ is a measure of the configurational disorder in the material, in direct analogy to the way in which the ambient temperature determines the intensity of low-energy fluctuations.  As such, $\chi_{ss}$ must be a function primarily of the plastic shear rate $\dot\epsilon^{pl}$, which determines the rate at which the atoms and dislocations are being caused to undergo rearrangements.  If this shear rate is so slow that the system has time to relax between rearrangement events, then the steady state of disorder is determined only by the number of rearrangements that have occurred and not by the rate at which they occurred.  That is, $\chi_{ss}$ must be some nonzero constant at shear rates appreciably smaller than atomic vibration frequencies.  It follows that $\rho_{ss}$ is independent of shear rate for relatively slow, steady-state deformations.  

However, the LMD simulations probe shear rates up to about $10^{10}/s$; thus, the constant-$\chi$ hypothesis cannot remain valid.   Determining a more general relation between the steady-state effective temperature $\chi_{ss}$ and the strain rate $\dot\epsilon^{pl}$ is a major theoretical challenge, possibly comparable to that of solving the glass problem.  My proposed compromise here is not the same as in our earlier analysis of strong-shock data in Ref. \cite{LBL-10}. There, my coauthors and I used a relation between strain rate and $\chi$ that Manning and I \cite{JSL-MANNING-TEFF-07} had deduced from the Haxton-Liu \cite{HAXTON-LIU-07} molecular-dynamics measurements of $\chi$ in a sheared glassy material.  Here I propose a much simpler way of describing this relation that does not invoke any glass-theoretic analogy. 

Before going into the details of this analysis, it will be useful to introduce some dimensionless notation. Let $\dot\epsilon^{pl} \equiv q/\tau_0$, where $\tau_0 = 10^{-12} s$ is an arbitrarily chosen atomic time scale.  Then, let 
 $\tilde\chi \equiv \chi/e_D$ and $\tilde\rho \equiv a^2\,\rho$, so that Eq.(\ref{rhoss}) becomes
\begin{equation}
\label{rhoss2}
\tilde\rho_{ss} = e^{-\,1/\tilde\chi_{ss}}.  
\end{equation}
The preceding discussion implies that $\tilde\chi_{ss}$ is a function only of $q$ and, moreover, that it is equal to a constant, say $\tilde\chi_0$, for strain rates appreciably smaller than atomic relaxation rates, i.e. for $q \ll 1$.  Using a variation of the Lindemann melting law, I estimated in \cite{LBL-10,JSL-17a} that $\tilde\chi_0 \cong 0.25$, which has turned out to be reasonably accurate.  I retain that estimate here in order that this high-rate analysis have some chance of making contact with experimental data at lower strain rates.     

To extrapolate from the small to the large strain-rate regimes, I postulate the following minimal formula for $\tilde\chi_{ss}(q)$:
\begin{equation}
\label{q-chi}
\tilde\chi_{ss}(q) =\tilde\chi_0 + {B\over [\ln(q_0/q)]^2}.
\end{equation}  
I have found this formula largely by trial and error, insisting that a small set of parameters, $B$ and $q_0$ in this case, be sufficient to describe the LMD data at six different strain rates for both the steady-state dislocation densities and the flow stresses.  Note that $q_0$ sets an upper limit for the strain rate, above which the system of dislocations effectively melts, perhaps via a sequence of twinning transitions as observed by LMD.  

Results of this analysis are shown in Fig. \ref{Fig1}, where I have plotted the dislocation density $\rho$ in units of $10^{17} /m^2$ as a function of the logarithm of the strain rate $\dot\epsilon^{pl}  = q/\tau_0$.  The data points are taken from LMD. As mentioned above, I set $\tilde\chi_0 = 0.25$; and I found that $B= 4.6$ and $q_0 = 0.008$.  I also found that $ 1/a^2 = 3.2 \times 10^{17}/m^2$. This means that $a \cong 1.8 \times 10^{-9} m$, which is bigger by a factor of about six than the atomic spacing reported in LMD.  $1/a^2$ is the dislocation density at which $q \to q_0$. 

\section {Basic Elements of the Thermodynamic Dislocation Theory: The Depinning Mechanism}

The key dynamical premise of the thermodynamic theory is that the dominant rate-controlling mechanism during rapid deformation is thermally activated depinning of entangled dislocations.  The depinning analysis starts with Orowan's relation between the plastic strain rate $\dot\epsilon^{pl}$, the dislocation density $\rho$, and the average dislocation velocity $v$:
\begin{equation}
\label{Orowan}
\dot\epsilon^{pl}= \rho\,b\,v,
\end{equation}
where $b \cong a$ is the magnitude of the Burgers vector.  If a depinned dislocation segment moves an average distance  $\ell = 1/\sqrt{\rho}$ between pinning sites, then $v = \ell/\tau_P$, where $1/\tau_P$ is a thermally activated depinning rate given by
\begin{equation}
\label{tauP}
{1\over \tau_P} = {1\over \tau_0}\,e^{- U_P(\sigma)/k_B T}.
\end{equation} 
 
The activation barrier $U_P(\sigma)$ must be a decreasing function of the stress $\sigma$. For dimensional reasons, $\sigma$ should be expressed here in units of some physically relevant stress, which we can identify as the Taylor stress $\sigma_T$ .  To see how this works, suppose that a pinned pair of dislocations must be separated by a distance $a' \ll a$ in order to break the bond between them. If these dislocations remain pinned to other dislocations at distances $\ell$,  then this displacement is equivalent to a strain of order $a'/\ell= a'\sqrt{\rho}$ and a corresponding stress of order $\mu\,a'\,\sqrt{\rho}$, where $\mu$ is the shear modulus.   Thus  
\begin{equation}
\label{sigmaT}
\sigma_T = \mu\,{a'\over \ell} \equiv \mu_T\,\sqrt{\tilde\rho};~~~\mu_T = (a'/a)\,\mu,
\end{equation}
where $\sigma_T$ is the Taylor stress, rederived here by an argument not very different from the one that Taylor used in his 1934 paper.\cite{TAYLOR-34} The factor $\mu_T$ also may contain a dimensionless factor of order unity to correct for uncertainty in the exponential function assumed in the following equation for $U_P(\sigma)$.  As in \cite{LBL-10}, write
\begin{equation}
\label{UP}
U_P(\sigma) = k_B\,T_P\,e^{- \sigma/\sigma_T(\tilde\rho)},
\end{equation}
where $k_B\,T_P$ is the pinning energy at zero stress.

The formula for the strain rate $q = \ell/\tau_P$ becomes 
\begin{equation}
\label{qdef}
q = \sqrt{\tilde\rho}\, \exp \Bigl[- {T_P\over T} e^{-\sigma/\sigma_T(\tilde\rho)}\Bigr].
\end{equation}
Here I have neglected a prefactor $b/a \cong 1$ that could, if necessary, have been absorbed into the time scale $\tau_0$. Now solve Eq.(\ref{qdef}) for $\sigma$ as a function of $\tilde\rho$, $q$ and  $T$. The result is:
\begin{equation}
\label{sigmadef}
\sigma = \sigma_T(\tilde\rho)\,\,\nu(\tilde\rho,q,T),
\end{equation} 
where 
\begin{equation}
\label{nudef}
\nu(\tilde\rho,q,T) = \ln\Bigl({T_P\over T}\Bigr) - \ln\Biggl[\ln\Bigl({\sqrt{\tilde\rho}\over q}\Bigr)\Biggr] .
\end{equation}
The quantity $\nu$ is a very slowly varying function of its arguments, consistent with the well known but approximate validity of the Taylor formula. 

\begin{figure}[h]
\begin{center}
\includegraphics[width=\linewidth] {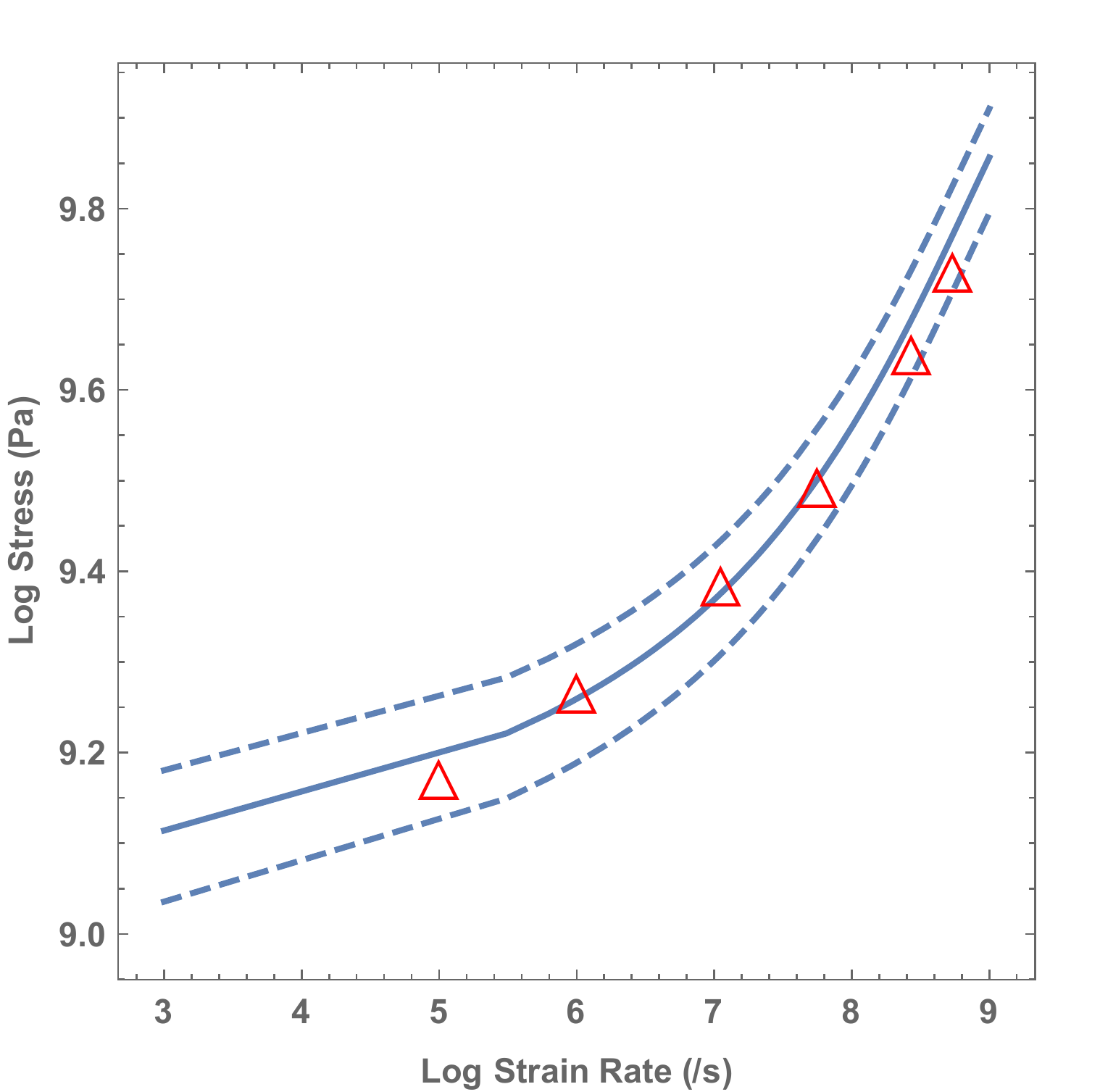}
\caption{Steady-state stress as a function of strain rate for temperatures $150\,K,\,300\,K,\, 600\,K$ from top to bottom. The data points at $300\,K$ are taken from \cite{Bulatov-17}}
\label{Fig2}
\end{center}
\end{figure}

Figure \ref{Fig2} shows six measured values of the steady-state stress as a function of strain rate at $T = 300\,K$, and my fit to these points using Eqs.(\ref{sigmadef}) and (\ref{nudef}).  With the parameters already determined from the analysis of the dislocation density shown in Fig.\ref{Fig1}, I need only two more parameters to compute the stress curve.  Fitting the data tells me that $T/T_P = 0.0008$ and $\mu_T = 2\, GPa$.  Again, this curve shows the transition between low-strain-rate and high-strain-rate behaviors predicted by the effective-temperature analysis.  In the conventional literature, this transition is often said to be evidence for increased phonon drag at high strain rates, and is often represented by an additive term in expressions for the stress. So far as I know, this interpretation has no basis in the LMD simulations or in any laboratory experiments.  The present theory says simply that higher strain rates produce higher effective disorder temperatures, and therefore produce the increased dislocation densities shown in Fig.\ref{Fig1} and the higher stresses in Fig.\ref{Fig2} .

In contrast to the dislocation-density analysis, the parameter $T/T_P$  in the stress analysis is explicitly temperature dependent, and $\mu_T$ is proportional to the elastic shear modulus $\mu$ which is known to depend on $T$.  To emphasize this point, I have plotted two extra curves in Fig.\ref{Fig2}, the lower curve at $T = 600\,K$ and the upper at $150\,K$, below and above the room-temperature curve at $300\,K$.  These two curves are plotted using only the explicit temperature dependence without assuming any temperature dependence of $\mu_T$, and are meant to be only illustrative rather than accurate predictions.

\section{Non-Steady-State Behaviors}
\label{stress-peak}

\begin{figure}[h]
\begin{center}
\includegraphics[width=\linewidth] {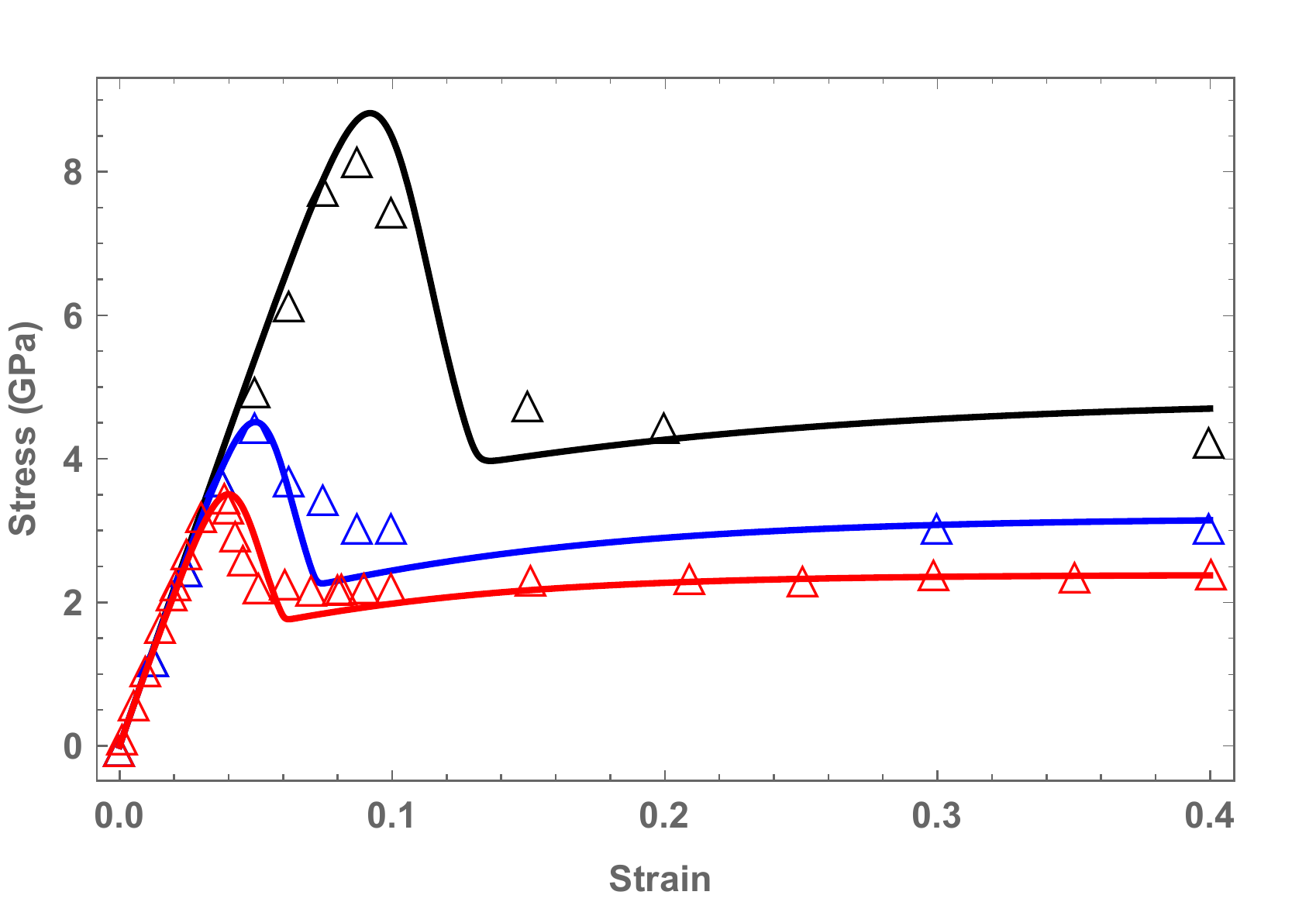}
\caption{Stresses as functions of strain for strain rates X1, X5, and X25.}
\label{Fig3}
\end{center}
\end{figure}

\begin{figure}[h]
\begin{center}
\includegraphics[width=\linewidth] {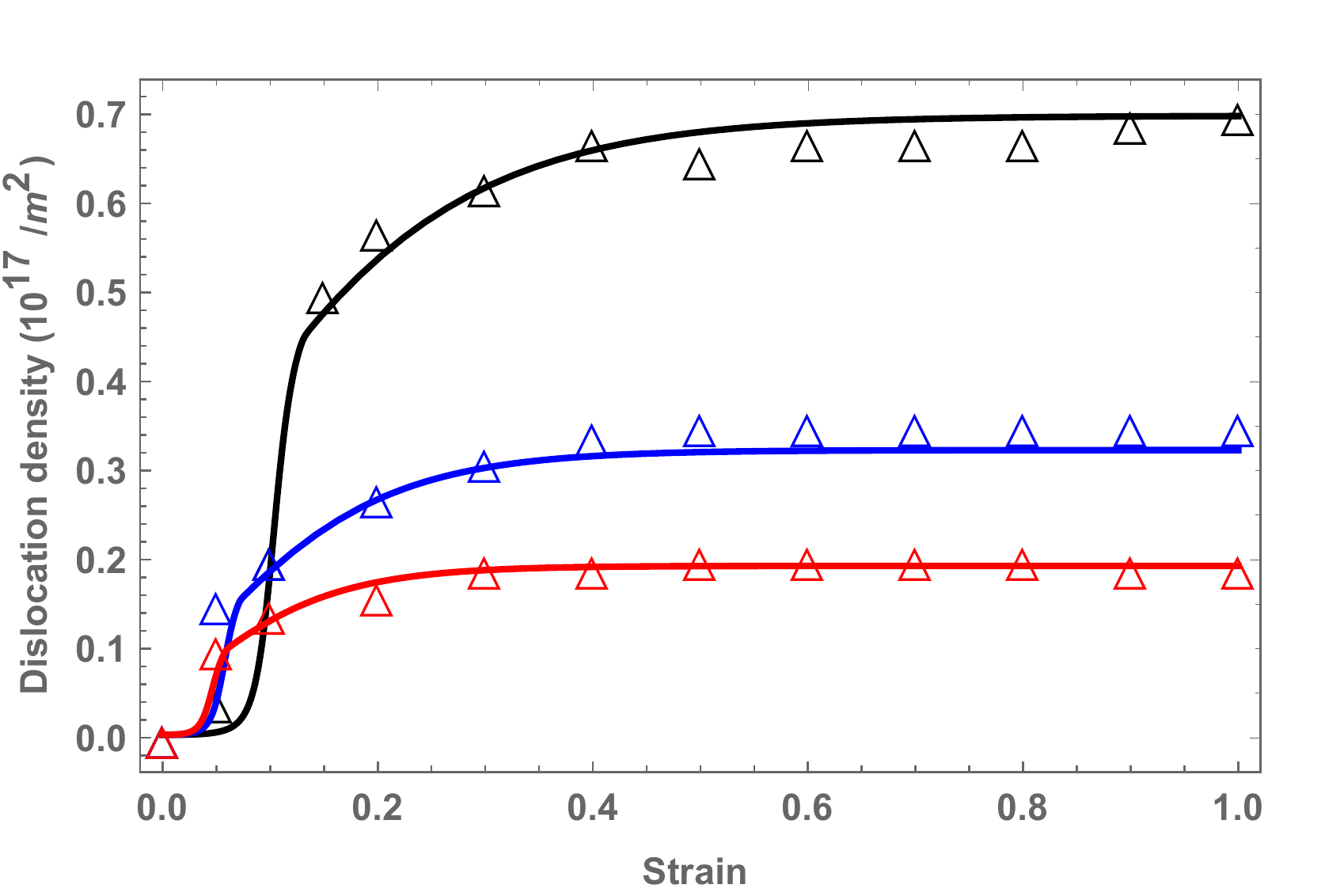}
\caption{Dislocation densities as functions of strain for strain rates X1, X5, and X25}
\label{Fig4}
\end{center}
\end{figure}

Turn now to the more unusual part of the LMD simulations -- the nonmonotonic behavior of the stress at the onset of plastic deformation.  The core of the LMD data is a set of three room-temperature, stress-strain curves at strain rates $\dot\epsilon = 1.1 \times 10^7\,s^{-1}$, $5.55 \times 10^7\,s^{-1}$, and $2.77 \times 10^8\,s^{-1}$, denoted by the authors as $X1$, $X5$, and $X25$ respectively. There are corresponding graphs of dislocation densities as functions of strain for each of the three strain rates.  These behaviors are shown by the data points here in Figs.   \ref{Fig3} and \ref{Fig4}.

I start by writing the relevant thermodynamic equations of motion. Because the LMD simulations all involve systems subject to time invariant  strain rates, say $\dot\epsilon \equiv Q/\tau_0$, it is convenient to replace the time $t$ by the total elastic plus plastic strain $\epsilon$, and to let $\tau_0\,\partial/\partial t \to Q\,\partial/\partial \epsilon$.  Then, assuming that the elastic and plastic strains are simply additive (``hypo-elasto-plasticity''), the equation of motion for the stress $\sigma(\epsilon)$ becomes
\begin{equation}
\label{sigmadot}
{\partial\sigma\over \partial\epsilon} = \mu_{eff}\,\Biggl[1 - {q(\epsilon)\over Q}\Biggr], 
\end{equation}
where $\mu_{eff}$ is the elastic modulus appropriate for the shear component of the stress in the LMD compressive geometry,  and $q(\epsilon)/\tau_0$ is the $\epsilon$-dependent plastic strain rate, which is yet to be determined as a function of $\sigma$.

The equation of motion for the dimensionless dislocation density $\tilde\rho(\epsilon)$ is
\begin{equation}
\label{dotrho}
{\partial\tilde\rho\over \partial\epsilon} = \kappa\,\,{\sigma(\epsilon)\,q(\epsilon)\over Q}\, \Biggl[1 - {\tilde\rho(\epsilon)\over \tilde\rho_{ss}}\Biggr],~~~\tilde\rho_{ss} = e^{- 1/\tilde\chi_{ss}}.
\end{equation}
This is a statement of energy conservation supplemented by a detailed-balance approximation.  $\kappa$ is an energy-conversion coefficient, proportional to the fraction of the input power $\sigma\,q$ that is converted into dislocations.  (Earlier versions of this equation used a more complicated prefactor that could be evaluated directly in terms of the early-stage hardening rate.  There is no need for that here.) The second term in square brackets in Eq.(\ref{dotrho}) is proportional to the annihilation rate, determined by the detailed-balance condition that $\tilde\rho \to \tilde\rho_{ss}$ in the limit of steady-state motion.  It will be simplest for present purposes to assume that $\tilde\chi_{ss}$ depends only on the total strain rate $Q$ and not on just the plastic part.  As will be seen in the next paragraph, effective-temperature dynamics seems (so far) to play no role in the onset behavior.

It has been assumed since the early work of Peierls and Nabarro that a dislocation in free motion through a crystal is subject to a drag force proportional to its velocity.  Conversely, its velocity is proportional to the driving stress.  At the extremely small densities of dislocations in the initial stages of the LMD simulations, it is no longer reasonable to assume that the time spent by a dislocation segment in moving from one pinning site to another is neglible compared to the pinning time; thus we must account for a drag time $\tau_D$ analogous to the pinning time $\tau_P$ introduced in Eq.(\ref{tauP}). 

The velocity $v$ appearing in the Orowan relation, Eq.(\ref{Orowan}), becomes 
\begin{equation}
v = {\ell\over \tau_P + \tau_D},
\end{equation}
where, as before, $\ell = a/\sqrt{\tilde\rho}$.  (Modifying this length scale by including a density of other obstacles seems not to make much difference.)  Let $\eta$ be a drag coefficient with dimensions of stress, so that $v_{drag} \equiv \ell/ \tau_D \equiv b\,\sigma/\eta\,\tau_0$.  The result is that Eq.(\ref{qdef}) becomes 
\begin{equation}
\label{qdef2}
q = f(\sigma, \tilde\rho)\,\Bigl[ 1 + {\eta\,f(\sigma,\tilde\rho)\over \sigma\,\tilde\rho}\Bigr]^{-1},
\end{equation}
where, as in Eq.(\ref{qdef}),
\begin{equation}
\label{fdef}
f(\sigma, \tilde\rho) = 
\sqrt{\tilde\rho}\,\exp \Bigl[- {T_P\over T} e^{-\sigma/\sigma_T(\tilde\rho)}\Bigr].
\end{equation}
Note that the drag correction, proportional to $\eta$, requires small values of $\tilde\rho$ in order to be non-negligible.  Note especially that the right-hand side of Eq.(\ref{qdef2}) is a highly nonlinear function of $\sigma$ and $\tilde\rho$.  It could not possibly be consistent with the conventional assumption that the stress is simply the  sum of drag and barrier-resistance terms. But it does satisfy the condition that, when $\eta$ becomes dominantly large, $q \to \tilde\rho\,\sigma/\eta$.

Both the LMD results and the solutions of these equations are shown in Figs. \ref{Fig3} and \ref{Fig4}.  The data points are taken directly from LMD. Note that I have chosen different strain scales for these graphs in order to focus on what seem to be the most important features -- the smaller strains for the stress and the larger ones for the more slowly relaxing dislocation density. In addition to the parameters obtained from the steady-state data given above, the parameters used for computing {\it all} six of these theoretical curves were: $\mu_{eff} = 110\,GPa$,  $\kappa = 0.35$, and $\eta = 1000$.  I also needed values for the steady-state effective temperatures obtained by solving Eq.(\ref{q-chi}) for $\tilde\chi_{ss}(q)$ at each $q$.  The latter values were $\tilde\chi_{ss} = 0.356,\, 0.436$, and $0.657$  for X1, X5, and X25 respectively.  The initial values of $\tilde\rho$, i.e. $\tilde\rho(\epsilon = 0)$, were all taken to be $10^{-3}$.

The agreement between the theory and the simulations seen in Figs. \ref{Fig3} and \ref{Fig4} tells us that the theory accounts satisfactorily for the crossover between drag and entanglement effects.  For example, note the dramatic increases in the dislocation densities in Fig.\ref{Fig4} at the points where the stresses are passing their peaks in Fig. \ref{Fig3}.  This behavior is closely related to the highly nonlinear features of the double-exponential function $f(\sigma, \tilde\rho)$ in Eq.(\ref{fdef}).  This is the same function that produces sharp yielding transitions when the initial dislocation densities are large \cite{JSL-17}, and also triggers adiabatic shear banding instabilities when coupled to changes in the ambient temperature. \cite{LTL-18}   Clearly, the agreement between theory and simulations is not perfect, especially at the stress peaks.  This agreement might be improved by allowing parameters such as $\eta$ and $\kappa$ to be rate dependent.  There may be some physical reason why the after-peak stresses drop faster and further in the theory than in the simulations; but uncertainties in the simulation data might also be playing a role.

\section{Concluding Remarks}

Even more than in earlier comparisons of the thermodynamic theory with experimental data, the agreement shown here between the theory and numerical simulations is remarkable.  I cannot overstate the importance of the simulations.  They force the theory to account for the dislocation densities as well as the stresses and thus lead to new insights -- Eq.(\ref{q-chi}) for example.  They lend credibility to the previous thermodynamic postulates, and they demonstrate that some elements of the conventional phenomenological analyses are incorrect. 

This theoretical success continues despite the fact that the thermodynamic theory seems drastically too simple.  It pays no attention at all to the huge amount of detailed observational information that has been accumulating for more than half a century about how dislocations move in various crystalline environments or how they interact with many different kinds of crystalline defects.  This body of information supposedly has been incorporated into the phenomenological literature; but I think that little if any of that literature can be considered reliable in view of its fundamental inconsistencies.  

At the same time, it seems obvious that we have a major opportunity to explore basic phenomena in new ways. For example, a  large part of the LMD paper is devoted to the authors' discovery that, at higher strain rates or lower temperatures than those discussed here, their simulated tantalum crystal undergoes extensive twinning transformations.  Might these be phase transitions driven by the increasing effective temperature?  Or is there yet some other fundamental process occurring?  More generally, it seems that a great many of the conventional phenomenological connections between microscopic observations (pile-ups, stacking faults, cellular dislocation patterns, etc.) and macroscopic plasticity need to be reexamined in the thermodynamic framework. There is much work that needs to be done. 

\begin{acknowledgments}

Vasily Bulatov has been extremely helpful in providing data and in explaining elements of the Livermore simulations.   I thank him greatly. I was supported in part by the U.S. Department of Energy, Office of Basic Energy Sciences, Materials Science and Engineering Division, DE-AC05-00OR-22725, through a subcontract from Oak Ridge National Laboratory.  

\end{acknowledgments}

\end{document}